\documentclass[conference]{IEEEtran}
\IEEEoverridecommandlockouts
\usepackage{cite}
\usepackage{amsmath,amssymb,amsfonts}
\usepackage{algorithmic}
\usepackage{graphicx}
\usepackage{textcomp}
\usepackage{xcolor}
\usepackage{booktabs}
\usepackage{caption}
\usepackage{hyperref}
\usepackage{amssymb}

\DeclareMathOperator*{\argmin}{arg\,min}
\usepackage{bm}
\usepackage{multicol, multirow}
\usepackage{comment}
\usepackage{tikz}
\usepackage{textcomp}
\usepackage{xpatch}

\usepackage{xstring}
\usepackage{collcell}

\newcommand*{\SuperScriptSameStyle}[1]{%
  \ensuremath{%
    \mathchoice
      {{}^{\displaystyle #1}}%
      {{}^{\textstyle #1}}%
      {{}^{\scriptstyle #1}}%
      {{}^{\scriptscriptstyle #1}}%
  }%
}

\newcommand*{\oneS}{\SuperScriptSameStyle{*}}
\newcommand*{\twoS}{\SuperScriptSameStyle{**}}
\newcommand*{\threeS}{\SuperScriptSameStyle{*{*}*}}

\def\L{{\cal L}}

\def\BibTeX{{\rm B\kern-.05em{\sc i\kern-.025em b}\kern-.08em
    T\kern-.1667em\lower.7ex\hbox{E}\kern-.125emX}}
\begin{document}

\title{Continual Learning for Monolingual End-to-End Automatic Speech Recognition}

\author{\IEEEauthorblockN{Steven Vander Eeckt and Hugo Van hamme} \\
\IEEEauthorblockA{KU Leuven \\ 
     Department Electrical Engineering ESAT-PSI\\
     Kasteelpark Arenberg 10, Bus 2441, B-3001 Leuven Belgium\\
     \textit{\{steven.vandereeckt, hugo.vanhamme\}@esat.kuleuven.be}}
}

\maketitle

\begin{abstract}
Adapting Automatic Speech Recognition (ASR) models to new domains results in a deterioration of performance on the original domain(s), a phenomenon called Catastrophic Forgetting (CF). Even monolingual ASR models cannot be extended to new accents, dialects, topics, etc. without suffering from CF, making them unable to be continually enhanced without storing all past data. Fortunately, Continual Learning (CL) methods, which aim to enable continual adaptation while overcoming CF, can be used. In this paper, we implement an extensive number of CL methods for End-to-End ASR and test and compare their ability to extend a monolingual Hybrid CTC-Transformer model across four new tasks. We find that the best performing CL method closes the gap between the fine-tuned model (lower bound) and the model trained jointly on all tasks (upper bound) by more than $\mathbf{40\%}$, while requiring access to only $\mathbf{0.6\%}$ of the original data. 
\end{abstract}


\section{Introduction}
Automatic Speech Recognition (ASR) has greatly progressed in recent years, moving from Hidden Markov Model (HMM) to End-to-End (E2E) models. However, like the Artificial Neural Networks (ANN) they use, E2E ASR models suffer from Catastrophic Forgetting (CF) \cite{catastrophicforgetting} when adapted to new tasks, even for monolingual tasks: it suffices that the data distributions of the old and new tasks differ for CF to occur. \\
Fortunately, many Continual Learning (CL) methods have been proposed in the image classification community, enabling ANNs to learn continually without suffering from CF. Following \cite{defy}, CL methods are categorized into three groups: i) the regularization-based methods use a regularization loss to train new tasks such that it does not hurt the performance of previous tasks. \cite{ewc, mas, si} do this by estimating the importance of each parameter to previous tasks and using these importance weights in a weighted L2 regularization to learn new ones. \cite{lwf} uses knowledge distillation \cite{knowledgedistillation} on the new tasks' data to transfer knowledge from the old to the new model; ii) the replay-based methods store a set of representative samples in a memory to rehearse old tasks when learning new ones. Most straightforward are \cite{er, ber, tem}, which train on the new task and the memory jointly. Alternatively, \cite{gem, agem} focus on gradient alignment between old and new tasks; iii) the architectural-based methods increase the model's capacity when learning new tasks. The latter are not considered in this paper. \\
Regarding ASR and, especially, E2E ASR, CL is a very new and unexplored topic. \cite{clacousticmodel, clhmmdnn} apply CL to the acoustic model of a HMM-based ASR model. \cite{cwav2vec} considers CL for the pre-trained wav2vec2 model \cite{wav2vec2}. \cite{synthasr} combines a Text-to-Speech and ASR model to prevent forgetting. \cite{fu2021incremental} applies Learning without Memorizing \cite{lwm} to E2E ASR, focusing on a scenario where subsequent tasks are much smaller than the initial one. Finally, \cite{lifelongasr} implements four existing CL methods for E2E ASR. Compared to \cite{lifelongasr}, we implement an extra seven CL methods for E2E ASR. We test and compare their ability to continually extend and enhance a monolingual E2E ASR by training it on new data. To make the experiments more realistic, we run the methods without assuming access to validation sets of previous tasks to optimize hyper-parameters. Since for many CL methods, both regularization-based and rehearsal-based, the weight of the regularization is a crucial hyper-parameter, we propose, based on \cite{defy}, a simple and efficient way to determine this weight.

\section{Continual Learning for E2E ASR}
\label{sec:exp}
We first elaborate on the considered E2E ASR model as well as on the objective of Continual Learning for E2E ASR. \\
\textbf{Model.} Our model is the Hybrid CTC/Transformer from \cite{hybrid_ctctransformer}. Its loss during training is computed as:
\begin{equation}
    \L(X, y; \theta) = c \cdot \L^{c}(X, y; \theta) + (1-c) \cdot \L^{d}(X, y; \theta)
    \label{eq:loss}
\end{equation}
where $\L^{c}(X, y; \theta)$ and $\L^{d}(X, y; \theta)$ are, respectively, the CTC and Decoder Cross-Entropy (CE) loss of the model with parameters $\theta$ on utterance $X$ with ground truth $y$. As in \cite{hybrid_ctctransformer}, the weight of CTC for training and decoding is $c=0.3$. No Language Model is used during decoding. The outputs of the model are 300 word pieces, generated by the Sentence Piece model \cite{sentencepiece} on the training data of the first task. \\
\textbf{Notation.} Denote $f^{c}(X; \theta) \in \mathbb{R}^{L \times o}$ and $f^{d}(X; \theta) \in \mathbb{R}^{W \times o}$ the CTC and Decoder output, respectively, of the model with parameters $\theta$, given utterance $X$, with $L$, $W$ and $o$ the utterance length, output length and number of word pieces. During training, $f^{d}(X; \theta)$ is conditioned on ground truth $y$. \\
\textbf{Problem formulation.} Let $D_1, D_2, .., D_T$ represent the labeled training datasets of the $T$ tasks. If $\theta^t$ are the model's parameters after learning $t$ tasks, then the objective of the CL methods is to learn tasks $1,.., T$ in sequence such that after $T$ tasks, $\theta^T$, adapted from $\theta^{T-1}$, satisfies:
\begin{equation}
    \theta^T = \argmin_\theta \sum_{t=1}^T \sum_{(X,y) \in D_t} \L(X, y; \theta)
    \label{eq:obj}
\end{equation}
However, when learning task $T$ on $D_T$, access to $D_1, ..., D_{T-1}$ is assumed to be lost (though storing a small number of utterances per task in a memory is allowed for the rehearsal-based methods), thus $\theta^T$ cannot be directly computed from \eqref{eq:obj}. In addition, we assume that we can no longer use the previous tasks' validation sets (to optimize hyper-parameters of the CL methods). We consider this a more realistic scenario.
\section{Continual Learning Methods}
\label{sec:methods}
We consider both regularization- and rehearsal-based methods. As E2E ASR models have a complex architecture and are computationally demanding to train, we focus on lightweight methods which are easily applicable to any ANN architecture, and have proven to work well in other domains. 
\subsection{Regularization-based Methods}
The regularization-based methods compute a regularization loss which is added to $\L(X, y; \theta)$ from \eqref{eq:loss} during training. \\
\textbf{Elastic Weight Consolidation (EWC).} After training task $t$, EWC \cite{ewc} computes the diagonal of the Fisher information matrix, denoted $\Omega^t$. $\Omega_{ii}^t$ is considered the importance weight of parameter $\theta_i$ for task $t$. Next, $\Omega^t$ is added to $\Omega^{\leq t}=\Omega^{\leq t-1} + \Omega^t$ as in \cite{ewc_more}, and used in the regularization loss to learn task $t+1$:
\begin{equation}
    \L_{ewc}(\theta) = \frac{\lambda}{2} (\theta-\theta^t)^T \Omega^{\leq t} (\theta-\theta^t)
    \label{eq:loss_ewc}
\end{equation}
Since $\Omega^t$ is diagonal, \eqref{eq:loss_ewc} reduces to a weighted L2 regularization, with weight $\Omega^t_{ii}$ for parameter $\theta_i$. \\
\textbf{Memory-Aware Synapses (MAS).} MAS \cite{mas} works similar as EWC, but computes (the diagonal of) $\Omega^t$ differently. Given that the ASR model has both a CTC and Decoder output, we compute $\Omega^t$ for MAS as follows: 
\begin{equation}
        \Omega_{ii}^t = \mathbb{E}_{X \sim D_t} \left[\left| c  \frac{\partial \|  f^{c}(X;\theta^t) \|^2}{\partial \theta_i} + 
        (1-c)  \frac{\partial \|  f_{}^{d}(X; \theta^t) \|^2}{\partial \theta_i} \right| \right]
\end{equation}
Next, the loss is exactly the same as for EWC in \eqref{eq:loss_ewc}. \\
\textbf{Continual learning with Sampled Quasi-Newton (CSQN).} CSQN \cite{csqn} was proposed to extend EWC by considering interactions between parameters. Starting from EWC's $\Omega^t$, CSQN considers quasi-Newton methods to compute low-rank approximations of the Hessian of the loss, which are then used as in \eqref{eq:loss_ewc} to regularize training. We consider both the standard and the reduced version, which was called BTREE in \cite{csqn} and which we here denote CSQN-BT. \\
\textbf{Learning Without Forgetting (LWF).} LWF \cite{lwf}, when learning task $t+1$, uses knowledge distillation \cite{knowledgedistillation} between the old model (with parameters $\theta^t$) as teacher and the current model (with parameters $\theta$) as student, on the new task's data:
\begin{equation}
    \begin{aligned}
    \L_{lwf} & (X; \theta) = \lambda \cdot (c \sum_{i=1}^L \sum_{j=1}^o \frac{f_{i,j}^{c}(X; \theta^t)}{\gamma} \log \frac{f_{i,j}^{c}(X; \theta)}{\gamma} \\
     & + (1 - c) \sum_{i=1}^W \sum_{j=1}^o \frac{f_{i,j}^{d}(X; \theta^t)}{\gamma} \log \frac{f_{i,j}^{d}(X;\theta)}{\gamma} )
    \end{aligned}
    \label{eq:loss_lwf}
\end{equation}
With $\gamma$ called the temperature. In our experiments, $\gamma=1$. \\
\subsection{Rehearsal-based Methods}
\label{subsec:reh}
The rehearsal-based methods use a small memory of exemplars of previous tasks to enable CL. \\
\textbf{Experience Replay (ER).} We consider three variants of ER \cite{er}. In the standard variant, the mini-batch from the current task is augmented with a mini-batch sampled from memory and sent through the model to compute the loss. As this may result in overfitting on the memory, the loss of the mini-batch sampled from memory may be given a weight $\lambda \in (0, 1)$, denoted ER ($\lambda$). Alternatively, as in \cite{ber}, the training set and memory can be merged to train on the resulting set, referred to as BER (Batch-level ER). \\
\textbf{Average-Gradient Episodic Memory (A-GEM)}. Consider $g=\frac{\partial \L(X, y; \theta)}{\partial \theta}$ with $(X, y)$ a mini-batch from the current task. Before $g$ is used to update the model, A-GEM \cite{agem} samples a mini-batch $(\tilde{X}, \tilde{y})$ from memory, and computes $g_{ref} = \frac{\partial \L(\tilde{X}, \tilde{y}; \theta)}{\partial \theta}$. If $g$ and $g_{ref}$ interfere, i.e. if $g^T g_{ref} < 0$, it updates $g$ with $g\gets g - \frac{g^T g_{ref}}{g_{ref}^T g_{ref}}g_{ref}$ such that the gradients align. The resulting gradient is used to update the model. A-GEM is the more efficient version of GEM (Gradient-Episodic Memory) \cite{gem}, which was the best method in \cite{lifelongasr}. \\
\textbf{Knowledge Distillation (KD).} KD uses the same loss as LWF in \eqref{eq:loss_lwf}, not computed on a mini-batch of the new task, but on a mini-batch sampled from the memory. Note that this loss is added to \eqref{eq:loss}, so the new task is still learned using the CE loss.

\section{Experiments}
Experiments were done in ESPnet \cite{watanabe2018espnet}. For detailed information and more extensive results, see \href{https://github.com/StevenVdEeckt/CGN\_CL\_Dialect}{our repository} \footnote{\url{https://github.com/StevenVdEeckt/CGN\_CL\_Dialect}}. \\
\textbf{Data.} We use the Corpus Gesproken Nederlands (CGN) dataset \cite{cgn}, which contains 900 hours of Dutch speech from both the Netherlands (NL) and Belgium (VL). We consider all except the more spontaneous speech and, based on the dialect of the speakers, split the data into four tasks: \textit{NL-main}, \textit{VL-main}, \textit{NL-rest}, \textit{VL-rest} (learned in this order). Each task is further split into a training, validation and test set. \\ 
\textbf{Training.} We use the optimizer from \cite{hybrid_ctctransformer}, with a learning rate of 10.0 for the first task and 1.0 for subsequent tasks. We allow models to run for 230 epochs, but stop early when the Token Error Rate (TER) at word piece level on the new task's validation set has not improved for 10 epochs. As in \cite{transformer}, we average the last 10 snapshots to obtain a final model. \\
\textbf{Determining $\bm{\lambda}$.} Many of the CL methods require setting a hyper-parameter $\lambda$, the weight of the regularization. Based on \cite{defy}, we propose a simple and efficient way to determine $\lambda$ for E2E ASR. First, we consider $\tau^{init}$, the TER (on the new task's validation set) of the initial model. Next, we adapt the model for five epochs without regularization, and compute its TER, obtaining $\tau^{no\_reg}$. Then, we set $\lambda$ to a high value and run the model for five epochs with regularization with weight $\lambda$. We compute the TER and obtain $\tau$. If $(\tau-\tau^{init}) / (\tau^{no\_reg} - \tau^{init}) > a$, i.e. if the gap between $\tau^{init}$ and $\tau^{no\_reg}$ is closed for at least $100a\%$, we return $\lambda$; else, we set $\lambda \gets p \lambda$ with $p \in (0, 1)$ and repeat the process. As such, determining $\lambda$ is done in a fast and efficient way and does not require access to a validation set of previous tasks. We determine $\lambda$ only for the first adaptation and then fix it. In our experiments, we set $a=0.85$ and $p=0.10$. Moreover, for each method, the initial value of $\lambda$ is a power of 10. \\
\textbf{Memory.} After learning a task, we sample 500 utterances from the training set to add to the memory. While sampling uniformly, we only consider utterances whose output length (i.e. number of word pieces in output) exceeds $0.40 \cdot mean\_length$, where $mean\_length$ is the average output length of the utterances in the training set, to make sure that all of the 500 utterances contain meaningful sentences. \\
\textbf{Baselines.} Following baselines are considered: (i) Fine-Tuning (FT): the model is adapted without CL method (lower bound); (ii) Joint (JT): trained from scratch on all tasks jointly; (iii) Continued Joint (CJT): adapted from previous task and trained on current and previous tasks jointly (upper bound). \\
\textbf{Metrics.} For each method, we report the Average WER (AWER), Backward Transfer (BWT) and Forward Transfer (FWT) \cite{gem}, and Coverage (COV) \cite{clacousticmodel}. Assuming $T$ tasks have been learned and $R_{i,j}$ is the WER on task $j$ after learning up to task $i$, $\text{AWER}=\sum_{i=1}^T R_{T,i}$, while the BWT is:
\begin{equation}
    \text{BWT} = \frac{1}{T-1} \sum_{i=1}^{T-1} -(R_{T,i} - R_{i,i})
\end{equation}
Note that using this definition, negative BWT indicates forgetting. Furthermore, we define FWT as: 
\begin{equation}
    \text{FWT} = \frac{1}{T-1} \sum_{i=2}^{T} -(R_{i,i} - R^{FT}_{i,i})
\end{equation}
where $R^{FT}_{i,j}$ is the WER on task $j$ after learning up to task $i$ with FT. FWT measures to which extent the model can exploit previously acquired knowledge to learn new tasks better. Positive FWT indicates better learning than FT. Finally, COV measures the extent to which the given method closes the gap between FT (lower bound) and CJT (upper bound) in terms of AWER. It is $0\%$ when the method performs as poor as FT, and $100\%$ when the method performs as well as CJT. In addition to AWER, BWT, FWT and COV, we report the storage requirements (Storage), expressed in an equivalent number of models (one model requiring 105 MB). \\
\textbf{Statistical significance.} We use the Wilcoxon signed-rank test on the number of errors per utterance \cite{Strik2000ComparingTR} to test the significance of the results, considering significance levels $\alpha=0.05$ ($\oneS$), $\alpha=0.01$ ($\twoS$) and $\alpha=0.001$ ($\threeS$).

\section{Results}
\label{sec:typestyle}
Table \ref{tab:final_results} shows the results after learning the four tasks in sequence. 
\begin{table}
    \centering
    \caption{Results after learning the four tasks. Significance level refers to improvement over baseline FT.}
    \begin{tabular}{l r r r r r }
    Model & AWER$\downarrow$\phantom{\threeS} & BWT$\uparrow$ & FWT$\uparrow$ & COV$\uparrow$ & Storage \\
    \toprule
    JT & 21.3\phantom{\threeS} & - & - & - & 260.54  \\
    CJT & 21.9\phantom{\threeS} & +2.5 & +0.5 & 100.0$\%$ & 261.54 \\
    \midrule
    FT & 27.3\phantom{\threeS} & -4.2 & - & 0.0$\%$ & 1.00 \\
    \midrule
    EWC & 28.3\phantom{\threeS} & \textbf{-0.7} & -4.8 & -18.9$\%$ & 2.00  \\
    MAS & 28.3\phantom{\threeS} & -1.1 & -4.4 & -18.9$\%$ & 2.00  \\
    CSQN & 27.7\phantom{\threeS} & -1.5 & -3.2 & -8.7$\%$ & 32.00 \\
    CSQN-BT & 27.8\phantom{\threeS} & -1.7 & -3.2 & -9.8$\%$ & 22.00  \\
    LWF & 26.6$\threeS$ & -3.3 & \textbf{+0.1} & 12.4$\%$ & 1.00 \\
    \midrule
    A-GEM & 26.1$\threeS$ & -2.5 & -0.0 & 22.0$\%$ & 3.24  \\
    ER & 28.0\phantom{\threeS} & -3.4 & -1.7 & -13.1$\%$ & 3.24 \\
    ER ($\lambda$) & 25.8$\threeS$ & -1.9 & -0.3 & 27.2$\%$ & 3.24 \\
    BER  & 26.4$\threeS$ & -2.8 & -0.2 & 16.7$\%$ & 3.24 \\
    KD & \textbf{25.0}$\threeS$ & -1.2 & +0.0 & \textbf{41.7$\%$} & 3.24 \\
    \bottomrule
    \end{tabular}
    \label{tab:final_results}
\end{table}
\begin{table}
    \centering
    \caption{WER on Test set and Memory of initial task \textit{NL-main} after the first adaptation to \textit{VL-main}. 'Initial' is the model trained on \textit{NL-main}, from which the other models are adapted.}
    \begin{tabular}{l r r}
    Model & Test & Memory \\
    \toprule 
    Initial & 27.1 & 10.7 \\
    FT & 33.0 & 21.8 \\
    \midrule 
    A-GEM & 31.0 & 2.6 \\
    ER & 32.7 & 0.0 \\
    ER ($\lambda$) & 30.6 & 0.4 \\
    BER & 32.3 & 13.9 \\
    KD & 29.4 & 10.5 \\
    \bottomrule
    \end{tabular}
    \label{tab:first_adapt_memory}
\end{table}
First, we note that FT indeed suffers from CF, while both JT and CJT are able learn the tasks well, with the latter reaching a positive BWT and FWT. \\
Considering the regularization-based methods, we find that the methods estimating which parameters are important experience difficulties learning the four tasks. This is especially true for EWC and MAS, both performing worse than FT. While CSQN and CSQN-BT, by considering interactions between parameters, perform slightly better, they still underperform FT. We hypothesize that the poor performance of these methods is due to the tasks (being very similar) having the same important parameters, which gives the model two options: either it updates these parameters, resulting in CF of previous tasks; or it leaves them unchanged, resulting in poor learning of the new tasks. In this experiment, EWC, MAS and CSQN reduce forgetting, but fail to learn the new tasks well. While EWC achieves the best BWT of all methods, it also attains the worst FWT. Note that the performance of EWC is in line with \cite{lifelongasr}, which also found EWC underperforming FT. Compared to EWC, MAS and CSQN, LWF performs much (and significantly, with $\alpha=0.001$) better, achieving the highest FWT (higher than FT). However, its COV is only $12.4\%$, as it reduces FT's forgetting (BWT) by only $21\%$. \\
Comparing LWF to KD, which uses the same regularization but computed on the memory instead of on the new task's data, we find that having access to a memory, even though it is only $0.6\%$ of the original data when learning the fourth task, yields big improvements (with significance level $\alpha=0.001$). KD attains a COV of over $40\%$, and learns the new tasks as well as FT, while reducing the latter's forgetting by more than $70\%$. It outperforms the other rehearsal-based methods by a large margin (with significance level $\alpha=0.001$). A-GEM, while it learns the new tasks well, still suffers from severe forgetting, reaching a COV of $22\%$. This is again consistent with \cite{lifelongasr}, which found GEM (of which A-GEM is a more efficient variant) outperforming LWF, while both improved the performance of FT.  Finally, ER performs worse than FT, as it suffers from CF and is unable to learn the new task well. Both BER, and especially ER ($\lambda$), perform much better, reaching a COV of $16.7\%$ and $27.2\%$, respectively. \\
Table \ref{tab:first_adapt_memory} gives us more insight into how the rehearsal-based methods work. It shows the WER on the memory and test set of the initial task \textit{NL-main} of models adapted to \textit{VL-main}. For ER, we note that it memorizes the memory completely, achieving 0.0 WER, and this generalizes very poorly to the test set. ER ($\lambda$) alleviates this, though it still almost perfectly memorizes the memory set. A-GEM, too, has a very low WER on the memory set. On the other hand, KD only very slightly improves on the memory set, but it is able to extract much more 'general' knowledge from it, limiting the forgetting on the test set much better than ER, ER ($\lambda$) or A-GEM.
\begin{figure}
    \centering
    \includegraphics[width=1\linewidth]{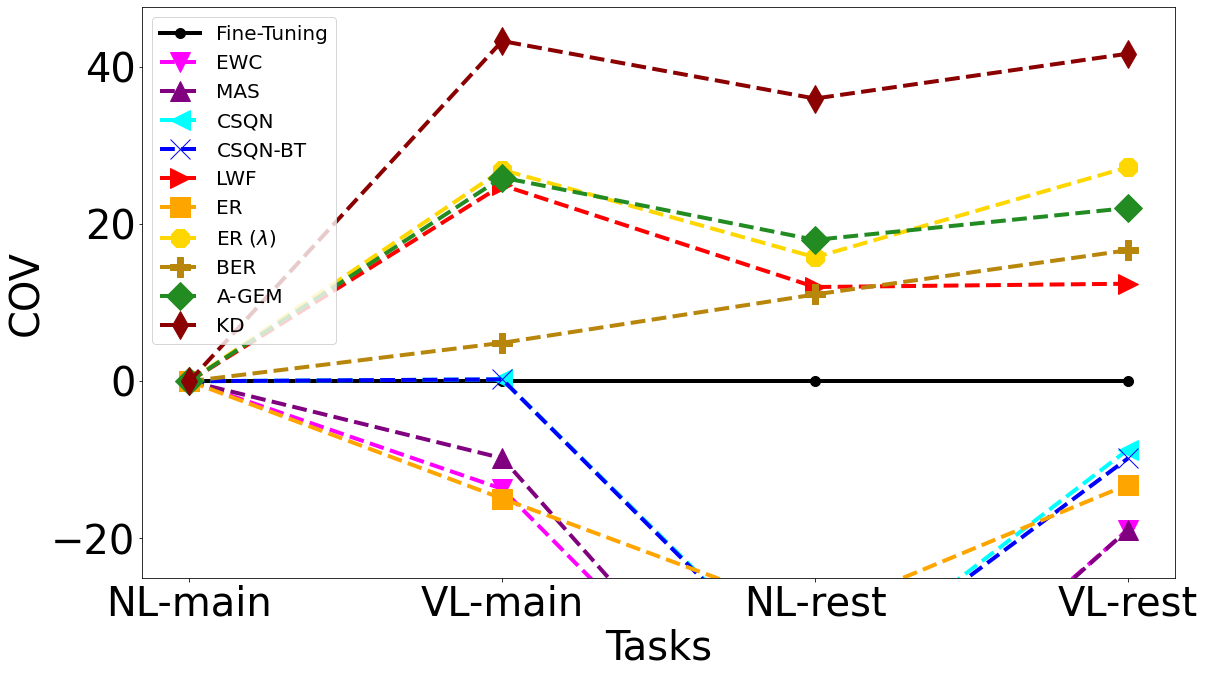}
    \caption{COV after learning the four tasks.}
    \label{fig:cov_final}
\end{figure} \\
Figure \ref{fig:cov_final} shows the COV after learning each task. We find that after two tasks, LWF performs as well as A-GEM and ER ($\lambda$). However, as more tasks are added, the gap between LWF, and A-GEM and ER ($\lambda)$ widens. Moreover, while BER after two tasks only slightly outperforms FT, its performance, relative to FT and the other CL methods, improves as more tasks are added. This is as expected, since BER, learning on the merger of the training set and the memory, clearly benefits from having a larger memory. Finally, note how many CL methods' performance drops when learning \textit{NL-rest}. That is because \textit{NL-main} and \textit{NL-rest} are very similar, so the CL methods should enable the model to learn \textit{NL-rest}, as this will also benefit \textit{NL-main}, while protecting \textit{VL-main}. While this is a realistic scenario when extending a monolingual ASR model, it turns out to be a very challenging one, especially for EWC, MAS and CSQN, which, by protecting \textit{NL-main}'s important parameters, are unable to exploit \textit{NL-rest} to further improve these parameters. 
\subsection{Increasing vs. Fixed Memory} 
The rehearsal-based methods from Table \ref{tab:final_results} had access to a memory with 500 utterances per task. In practice, it may be more desirable and/or feasible to have a memory with fixed size, especially as the number of tasks becomes large. To this end, we fix the size of the memory at 500. Table \ref{tab:fixed_memory} shows the results for A-GEM, ER ($\lambda$) and KD. 
\begin{table}
    \centering
    \caption{Results after learning the four tasks with fixed memory. Significance level refers to deterioration compared to corresponding method from Table \ref{tab:final_results} (with increasing memory). }
    \begin{tabular}{l l r r r r}
    Model & AWER$\downarrow$ & BWT$\uparrow$ & FWT$\uparrow$ & COV$\uparrow$ & Storage  \\
    \toprule
    A-GEM & 26.2$\oneS$ & -2.8 & -0.0 & 18.9$\%$ & 1.72 \\
    ER ($\lambda$) & 25.6 & -1.6 & -0.4 & 29.8$\%$ & 1.72 \\
    KD & \textbf{25.2}$\oneS$ & \textbf{-1.5} & \textbf{+0.1} & \textbf{38.3}$\%$ & 1.72 \\
    \bottomrule
    \end{tabular}
    \label{tab:fixed_memory}
\end{table} \\
We find that the differences with Table \ref{tab:final_results} are small. For A-GEM and KD, we observe a slight (though significant, with $\alpha=0.05$) deterioration of AWER. For ER ($\lambda$), on the other hand, there is even a minor improvement, which, not being statistically significant, is attributed due to chance. Even for a very small and fixed memory (with only $0.2\%$ of original training data after four tasks), A-GEM and, in particular, ER ($\lambda$) and KD are thus highly effective in enabling CL.
\subsection{Storage Requirements}
Table \ref{tab:final_results} shows the storage requirements of the CL methods to learn the fourth task. For all methods except JT, which for each new task starts from scratch, this requires, first of all, the storage of the model itself. In addition, the rehearsal-based methods require storing an equivalent of 2.24 models, as they need to store the utterances in the memory. \\
Compared to the rehearsal-based methods, the regularization-based methods are more storage efficient (in addition to not requiring data from previous tasks to be stored in a memory, which may be, due to e.g. privacy concerns, not always allowed). While LWF requires storing only the previous model, EWC and MAS, in addition, need to store the importance weights. Compared to the latter, CSQN and CSQN-BT are less storage efficient, due to the Hessian approximations.  \\
With an increasing memory, as in Table \ref{tab:final_results}, the rehearsal-based methods' storage requirements also increase linearly with the number of tasks. However, as we saw in Table \ref{tab:fixed_memory}, this can be overcome by fixing the memory size, with only a negligible deterioration in performance, enabling A-GEM, and, especially, ER ($\lambda$) and KD, to achieve excellent performance while being very storage efficient, requiring to store an equivalent of only 1.72 models (independent of number of tasks). Finally, note how JT and CJT, needing access to all data the model was ever trained on, require storing an equivalent of 260.54 and 261.54 models, respectively, making them clearly not a practical solution to overcome CF.
\section{Conclusion}
\label{sec:majhead}
In this paper, we implemented an extensive number of CL methods, and tested and compared their ability to extend a monolingual E2E ASR model across four tasks. Having access to a memory, though very small compared to the original training set, proved to be very beneficial, as the rehearsal-based methods generally performed much better than the regularization-based methods. To assure the former's storage requirements do not increase with the number of tasks, the memory size can be fixed with only a negligible degradation in performance. In general, thus, the rehearsal-based methods seem the best and most practical way to currently overcome CF in monolingual E2E ASR models; in particular KD, which closes the gap between the Fine-Tuned (lower bound) and Continued Joint model (upper bound) for $41.7\%$ and $38.3\%$ while having access to only $0.6\%$ and $0.2\%$, respectively, of the original data. In case storing utterances from previous tasks is not allowed, LWF seems to be the best option, as the other regularization-based methods, which have higher storage requirements, were unable to improve the Fine-Tuning lower bound. However, even in case only a very small number of utterances per task can be stored, it is advised to do so.

\bibliographystyle{IEEEbib}
\bibliography{conference_101719}

\end{document}